\documentclass[10pt,conference,hidelinks]{IEEEtran}

\usepackage{url}
\usepackage{subfig}
\usepackage{graphicx}
\usepackage{amsmath,amssymb}
\usepackage{hyperref}
\usepackage{listings}
\usepackage{tabularx}
\usepackage{xcolor}
\usepackage{comment}
\usepackage{listings,newtxtt}
\usepackage{tikz}
\usepackage{xcolor}

\usepackage{pbox}
\usepackage{comment}
\usepackage{csvsimple}
\usepackage{fancyhdr}
\usepackage{mathtools}
\usepackage{caption}
\usepackage{amsmath}
\usepackage{makecell}
\usepackage{tikz}

\usepackage{siunitx}
\begin{document}

\newcommand{\gem}{gem{5}}

\title{A Cycle-level Unified DRAM Cache Controller Model for 3DXPoint Memory Systems in \gem{}}

\author{
\\\IEEEauthorblockN{Maryam Babaie, Ayaz Akram, Jason Lowe-Power}
\IEEEauthorblockA{Department of Computer Science,
University of California, Davis\\
Email: \{mbabaie, yazakram, jlowepower\}@ucdavis.edu}}

\maketitle

\begin{abstract}

    To accommodate the growing memory footprints of today's applications, CPU vendors have employed large DRAM caches,
    backed by large non-volatile memories like Intel Optane (e.g., Intel's Cascade Lake).
    The existing computer architecture simulators do not provide support to
    model and evaluate systems which use DRAM devices as a cache to the non-volatile main memory.
    In this work, we present a cycle-level DRAM cache model which is integrated with \gem{}.
    This model leverages the flexibility of \gem{}'s memory devices models and full system support to enable exploration of many different DRAM cache designs.
    We demonstrate the usefulness of this new tool by exploring the design space of a DRAM cache controller through several case studies including the impact of
    scheduling policies, required buffering, combining different memory technologies (e.g., HBM, DDR3/4/5, 3DXPoint, High latency)
    as the cache and main memory, and the effect of wear-leveling when DRAM cache is backed by NVM main memory.
    We also perform experiments with real workloads in full-system simulations to validate the proposed model
    and show the sensitivity of these workloads to the DRAM cache sizes.

\end{abstract}


\section{Introduction}

The last decade has seen significant academic research on DRAM caches, and today
these ideas are becoming a reality with CPU vendors implementing DRAM cache-based computer systems, e.g., Intel's Cascade Lake and Sapphire Rapids.
Hardware-managed DRAM caches are seen as one way to enable heterogeneous memory systems (e.g., systems with DRAM and non-volatile memory) to be more easily programmable.
DRAM caches are transparent to the programmer and easier to use than manual data movement.

However, recent work has shown that these transparent hardware-based data movement designs are much less efficient than manual data movement~\cite{hildebrand2021case}.
While the work by Hildebrand et al.~\cite{hildebrand2021case} and other recent work investigating Intel's Cascade Lake systems provides some insight into real implementations on DRAM caches~\cite{izraelevitz2019basic,wang2020characterizing}, there is a gap in the community's access to cycle-level simulation models for DRAM caches.
This paper describes a new \gem{}-based model of a unified DRAM cache controller inspired by the Cascade Lake hardware to fill this gap.

Previous work has explored many aspects of DRAM cache design in simulation such as the replacement policy, caching granularity~\cite{qureshi2012fundamental,jevdjic2013stacked}, dram cache tag placement~\cite{huang2014atcache,loh2012supporting,loh2011efficiently}, associativity~\cite{qureshi2012fundamental,kotra2018chameleon,young2018accord}, and other metadata to improve performance~\cite{loh2011efficiently,jevdjic2013stacked,young2018accord}.
These mostly high-level memory system design investigations can appropriately be evaluated with trace-based or non-cycle-level simulation.
However, as shown in recent work, the micro-architecture of the unified DRAM and non-volatile main memory (NVRAM) controller can lead to unexpected performance pathologies not captured in these prior works (e.g., Hildebrand et al. showed that a dirty miss to the DRAM cache requires up to \emph{five accesses} to memory~\cite{hildebrand2021case}).

Thus, to better understand these realistic DRAM cache systems, it is imperative to build a detailed DRAM cache simulation model which can be used to perform a design space exploration around the DRAM cache idea.
The previous research works on DRAM cache design improvements do not provide any (open-source) DRAM cache modeling platform for a detailed micro-architectural and timing analysis.
To the best of our knowledge, most research works do not consider systems where the hardware-managed DRAM cache and NVRAM are sharing the same physical interface and are controlled by a unified memory controller (as is the case in real platforms like Intel Cascade Lake).

In this work, we describe our unified DRAM cache and main memory controller (\textit{UDCC}) cycle-level DRAM cache model for \gem{}~\cite{lowepower2020gem5}.
The protocol takes inspiration from the actual hardware providing DRAM cache, such as Intel's Cascade Lake, in which an NVRAM accompanies a DRAM cache as the off-chip main memory sharing the same bus.
To model such hardware, we leverage the cycle-level DRAM~\cite{hansson2014simulating} and NVRAM~\cite{gem5-workshop-presentation} models in \gem{}.
Our model implements the timing and micro-architectural details enforced by the memory interfaces including the DRAM timing constraints, scheduling policies, buffer sizes, and internal queues.
We propose a DRAM cache model that is direct-mapped, insert-on-miss, and write-back to model Intel's Cascade Lake design.

Using this model, we present validation data and investigate five case studies.

\emph{What is the impact of memory scheduling policies in a unified DRAM cache and memory controller?}
We find that using FR-FCFS is highly impactful when the cache hit ratio is high, but less so when the hit ratio is low and the NVRAM's bandwidth limits performance.

\emph{What is the impact of DRAM technology on performance and memory controller architecture?}
We find that higher performing memory technologies require more buffering to achieve peak performance. 
Moreover, we find that the composition of the memory access patterns and their hit/miss ratio on DRAM cache, 
can also affect the amount of buffering to achieve the peak bandwidth.

\emph{What is the impact of backing ``main memory'' performance?}
We find that while slower backing memory hurts performance, the performance of the backing memory does not have a significant affect on the micro-architecture of the cache controller.

\emph{What is the impact of the UDCC model for full-system applications?}
We find that our model shows similar performance characteristics on real applications as previously shown on real hardware providing further evidence for the importance of cycle-level simulation.

\emph{What is the impact of NVRAM wear leveling on memory system performance with a DRAM cache?}
We find that while wear leveling has a very small direct impact, the impact when using NVRAM as backing memory with a DRAM cache can be much higher. Although only 1 in 14,000 requests experience a wear-leveling event, the performance impact is up to an 8\% slowdown.

Our model is open-source and publicly available for the use of research community~\cite{dcacheGem5Code} and will be integrated into \gem{} mainstream. Using this new model which implements the micro-architectural details of realistic DRAM caches on a simulator can help find any potential improvement for the next generation of memory systems.


\section{Background}
\label{sec:background}

\subsection{Background on heterogeneous memory}

With the growing footprint of large scale applications, commercial products have been emerging in the market
to accommodate the memory requirements of the today's demanding workloads. Intel has introduced Optane Data-Center
Persistent-Memory-Modules (DC PMM) which is a non-volatile byte-addressable main memory (NVRAM) and can be a
replacement for conventional DRAM ~\cite{hildebrand2021case}. Even though NVRAM provides much larger capacity than DRAM, it has 3x longer latency and
60\% lower bandwidth than DRAM ~\cite{izraelevitz2019basic}. Hardware caching through a DRAM cache for NVRAM, has been employed in Intel's
Cascade Lake product to hide the long latency of NVRAM. Memory subsystem in Cascade Lake works in two modes,
1LM in which the NVRAM is persistent and directly accessible to the application, and 2LM in which DRAM DIMM
is the hardware cache for NVRAM. Recently, Intel announced the release of the new generation of Xeon processor called Sapphire Rapid~\cite{sapphire} which includes an HBM working in three modes,
HBM-Only, Flat Mode, and Cache Mode.
HBM-Only Mode considers HBM as the main memory of the system without
any DRAM memory. With the provided DDR5 in the system, HBM and DDR5 both work as the main memory in the
Flat Mode. Finally, HBM is used as a cache for DDR5 in the Cache Mode. To accommodate the research lines in this
domain, simulators must be able to correctly implement the real behavior of these memory technologies. Wang
et al.~\cite{wang2020characterizing} have studied the behavior of NVRAM in Intel's Cascade Lake and provided the
behavior seen in VANS model. However, they do not provide any support for the DRAM cache cooperating with NVRAM
in the 2LM mode. In this work we seek to model the DRAM cache protocol and implement it in gem5 simulator.

\subsection{Background on gem5's memory subsystem}

The \gem{} simulator is based on an event-driven simulation engine.
It supports models of many system components, including a memory controller, a detailed DRAM model, an NVM model, and a model for different CPUs, caches, and others.
The memory controller module added to \gem{} by Hansson et al.~\cite{hansson2014simulating} focuses on modeling the state transitions of the memory bus and the memory banks.
While it is not ``cycle accurate'', it is \emph{cycle level}, and the memory controller is designed to enable fast and accurate memory system exploration.
Hansson et al.~\cite{hansson2014simulating} showed that their memory controller model in \gem{} correlates well with a more detailed model DRAMSim2~\cite{rosenfeld2011dramsim2}, but leads to a 13\% reduction in simulation time on average.
The controller is responsible for address decoding (into a row, bank, column, and rank) and keeping track of the DRAM timings.
The memory controller is also configurable with different front-end and back-end latencies, which can model different memory controller designs and the impact of physical interfaces.

The memory controller in \gem{} was refactored recently~\cite{gem5-workshop-presentation}.
Now, it relies on two components to simulate a real memory controller. 
(1) The \textit{memory controller} receives commands from the CPU and enqueues them into appropriate queues and manages their scheduling to the memory device.
(2) The \textit{memory interface} deals with device/media-specific timings, manages the device-specific operations and communicates with the memory controller.

The unified DRAM cache memory controller (\textit{UDCC}) presented in this work implements a new memory controller module in \gem{} and requires only minor changes in the memory interface.
Like \gem's current memory controller, \textit{UDCC}'s goal is for cycle-level simulation to enable micro-architectural exploration and flexibility, not cycle-by-cycle accuracy to a single design.
\section{Unified DRAM cache controller}


\begin{figure}
  \centering
  \includegraphics[scale=0.45]{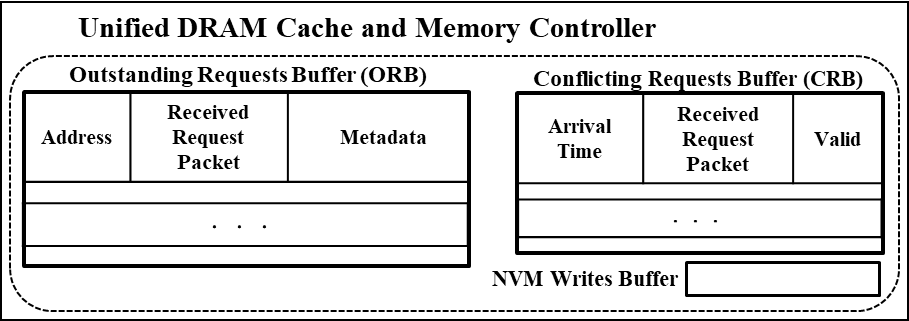}
  \vspace{-1ex}
  \caption{Hardware abstraction of \textit{UDCC}}
  \label{fig:buffers}
\end{figure}

\begin{figure}
  \centering
  \includegraphics[scale=0.4]{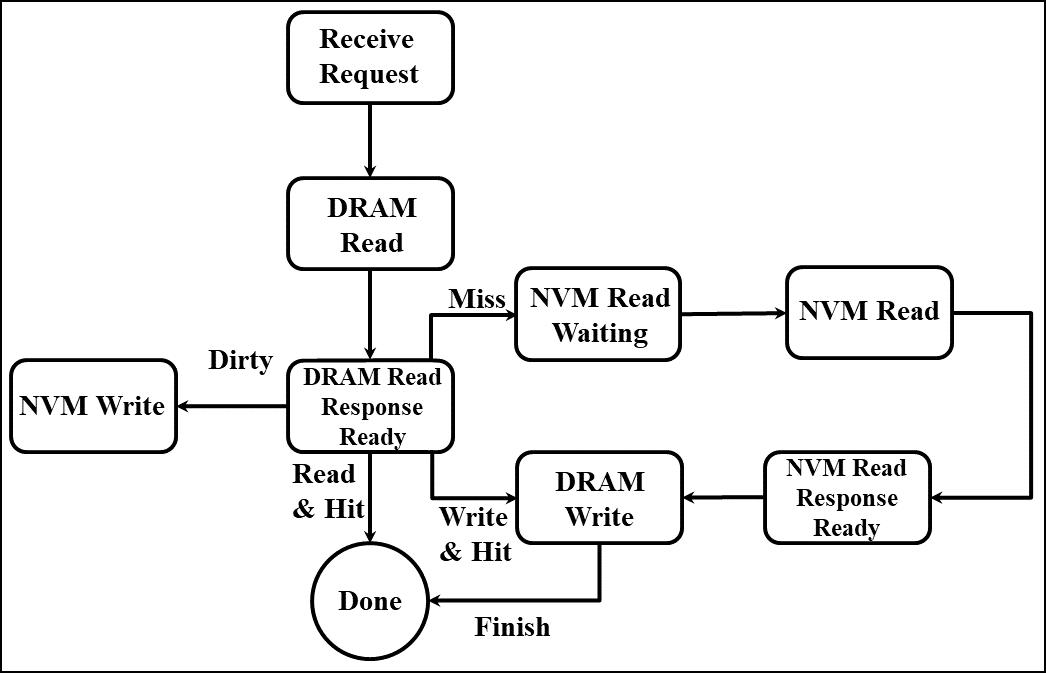}
  \vspace{-1ex}
  \caption{State machine followed by packets in \textit{UDCC}}
  \label{fig:StateMachine}
\end{figure}

We model a unified DRAM cache controller (\textit{UDCC}) in \gem{} to control a DRAM device (which acts as a cache of the main memory) and an NVM device (which serves as the main memory in the system).
\textit{UDCC} is largely based on Intel Cascade Lake's DRAM cache management strategy.
Figure~\ref{fig:buffers} provides a high-level overview of the modeled
controller's layout, and Figure~\ref{fig:StateMachine} shows the states that the memory packets transition through while residing in \textit{UDCC}.

We designed \textit{UDCC} to be a flexible model to explore different memory controller design parameters.
Thus, instead of modeling one specific micro-architecture with buffers for every device (e.g., separate DRAM and NVM read and write queues), we model the controller with a single large buffer called the \textit{Outstanding Request Buffer (ORB)}.
We also have a \textit{Conflict Request Buffer (CRB)} to store requests that are to the same cache line and must be serialized with current outstanding requests.
These buffers are shown in Figure~\ref{fig:buffers}.
All incoming memory requests reside in the \textit{ORB} unless a request conflicts with an already existing request in the \textit{ORB}.
Two requests are considered to be conflicting if they both map to the same DRAM cache block.
The conflicting request goes to the \textit{CRB} until the request it was conflicting with has been serviced and is taken out of the \textit{ORB}.
Each entry in these buffers contains other meta data in addition to the address of the request, as shown in Figure~\ref{fig:buffers}.
This metadata provides helpful information about the request, e.g., the memory request's current state and relative arrival time.
In addition to these two buffers, we also model an \textit{NVM WrBack} queue for the DRAM cache dirty lines that are to be written back to NVM main memory.

Figure~\ref{fig:StateMachine} presents the state machine followed by the memory packets while they are in \textit{UDCC}.
Since \gem{} is an event driven simulator, the simulation model relies on scheduled events to transition between these states.
To model the realistic buffers in the system, we can constrain the number of requests at one time in any state.
The states these memory packets can be in at any point are as follows:

\noindent
\textbf{Recv\_Pkt}: Every memory request received to the memory controller is in \textit{Recv\_Pkt} state initially.

\noindent
\textbf{DRAM\_Read:} Since every request must first check the tag in the cache, a memory packet moves to the \textit{DRAM\_Read} state when the time arrives that the DRAM can service this request. This transition is achieved by scheduling an event for the ready time of this packet. The device-specific memory interface provides the ready time. At this point, the request is also scheduled to the DRAM itself for a read.

\noindent
\textbf{DRAM\_Read\_Resp:} Once the response time of an already scheduled (to DRAM) packet arrives, the packet is moved to the \textit{DRAM\_Read\_Resp} state. At this point, the DRAM cache tags are checked to ascertain the hit/miss and clean/dirty status of the request and different actions are performed accordingly.

\noindent
\textbf{DRAM\_Write:} A memory packet can go to the \textit{DRAM\_Write} state in two scenarios: (1) if its tag read from DRAM cache is done (and found to be a hit) when it originally was a write packet, or (2) if we get a response from the NVM that is going to fill in DRAM cache (initiated by `a missing in DRAM cache' request).
Once in \textit{DRAM\_Write} state, packets are eventually written to the DRAM.

\noindent
\textbf{NVM\_Read\_Wait\_Issue:} Packets missing in DRAM cache (found out by a tag read in DRAM cache), are moved to this state so that they can eventually be scheduled to NVM.

\noindent
\textbf{NVM\_Read:} Once the ready time of packets in \textit{NVM\_Read\_Wait\_Issue} arrives (NVM interface is ready to accept the packet), they are scheduled to NVM and moved to \textit{NVM\_Read} state.

\noindent
\textbf{NVM\_Read\_Resp:} On getting a response back from the NVM device (when the NVM interface is done with the processing of the packet), the packet moves to \textit{NVM\_Read\_Resp} state where it will be eventually moved to the \textit{DRAM\_Write} state so that it can be written to the DRAM. If the packet was originally a read request (and missed in DRAM cache), at this point the response for this packet will also be sent to the requestor.

\noindent
\textbf{NVM\_Write:} The packets belonging to a dirty cache line in DRAM are moved to the \textit{NVM\_Write} state, so that they can be written to the main memory (NVM).

\noindent
\textbf{Done:} If a (read) packet was found to be a hit in the DRAM cache, the response will be sent back to the requestor and the packet will move to the \textit{Done} state.
Similarly, when a (write) packet is finally written to the DRAM cache, it moves to the \textit{Done} state. Packets in \textit{Done} state are removed from the \textit{ORB}.

Since we are modeling a memory-side cache that does not participate in the on-chip coherence protocol, we do not model the stored data in the cache.
We use the protocol described above to model the timing of the system, and for data we use the backing store of the backing memory (NVM in our case).
Similarly, while we model the timing of accesses to DRAM to check the tags, we functionally store the tags in a separate structure in our model. 
\section{Validation}
\label{sec:validation}

In this section, we present the validation of our DRAM cache model in \gem.
Any simulation model needs to be functionally correct, as well as provide accurate performance statistics.
To ensure functional validation,
we successfully performed Linux kernel boot tests in \gem{} full system mode, and
we ran some commonly used HPC class benchmarks in \gem{} full-system mode, which are presented in Section~\ref{sec:realBench} of this paper.

\begin{figure}
  \centering
  \includegraphics[scale=0.5]{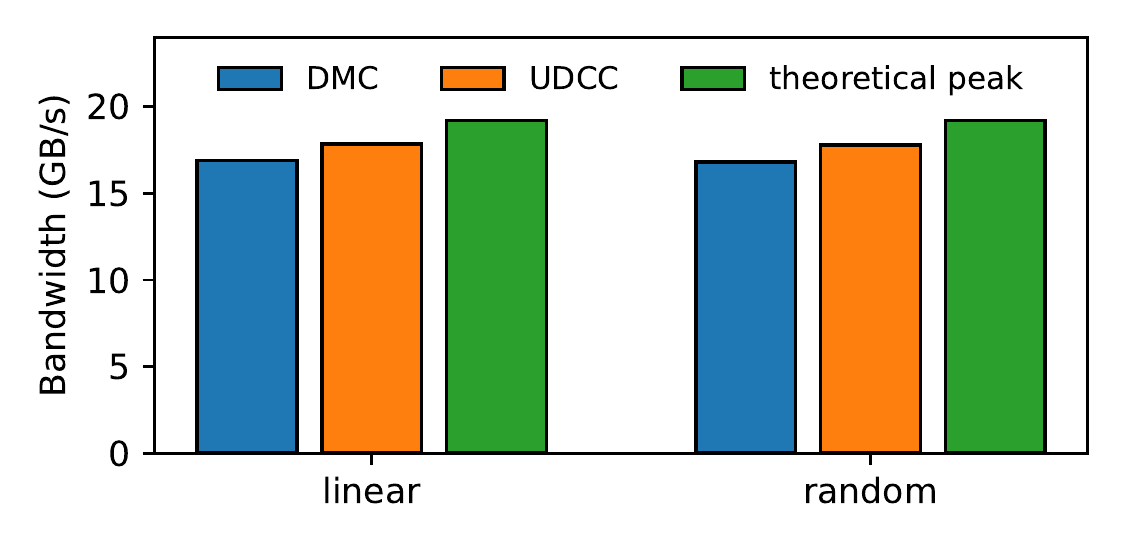}
  \vspace{-1ex}
  \caption{Read bandwidth comparison between \gem~default memory controller (\textit{DMC}), unified DRAM cache controller (\textit{UDCC}), and the theoretical peak bandwidth on DDR4. In case of \textit{UDCC}, all accesses hit in the DRAM cache. Two different traffic patterns (linear, random) are used. Read buffer size in \textit{MC}, and \textit{ORB} size in \textit{UDCC} is 256 entries.}
  \label{fig:valid}
\end{figure}

The evaluation of the accuracy of the performance model is generally not straightforward. Still, reasonable confidence can be established in the performance model by relying on the evaluation of performance numbers in controlled experiments.
We performed a comparison of the effective memory bandwidth observed with \gem's default memory controller (\textit{DMC}) and the unified DRAM cache controller (\textit{UDCC}).
We rely on a synthetic read memory traffic pattern such that (nearly) \textit{all requests will be hits in the DRAM cache} in the case of a system where we use \textit{UDCC}. 
We use both a linear and a random traffic pattern.
Figure~\ref{fig:valid} provides both controllers' read bandwidth and compares these numbers to the theoretical peak bandwidth possible with the DDR4 device used in this experiment.
As expected, we observe that the bandwidth attained by \textit{DMC} and \textit{UDCC} is very close.
The scheduling policy implemented in both controllers (details in Section~\ref{sec:sched}) can explain the slight difference. 
The effective bandwidth in both cases is less than the theoretical peak bandwidth.

\begin{figure}
  \centering
  \includegraphics[scale=0.5]{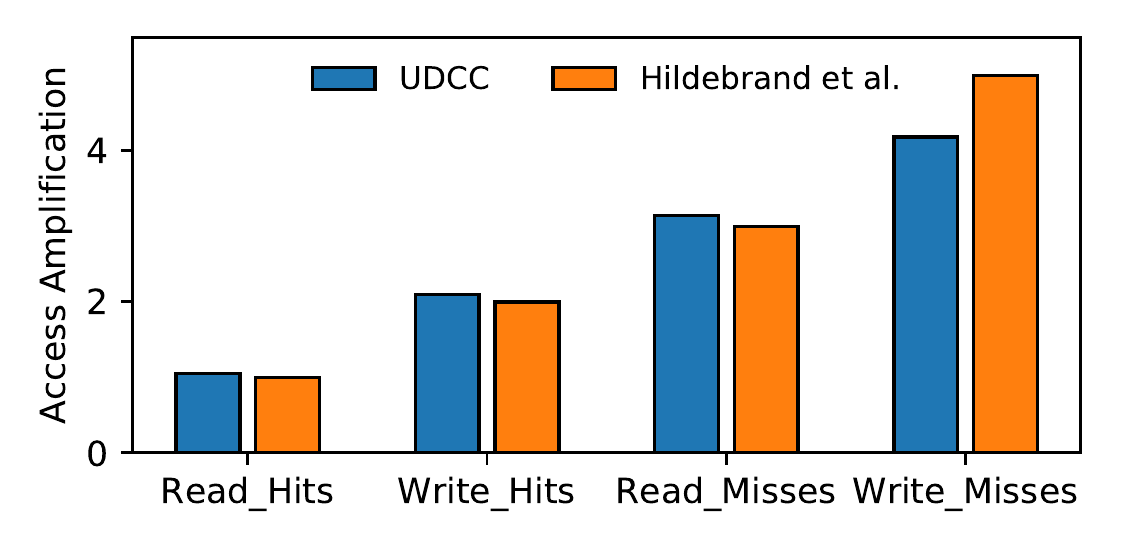}
  \vspace{-1ex}
  \caption{Access amplification observed with \textit{UDCC} in \gem{} and on the real hardware (Hildebrand et al.~\cite{hildebrand2021case}).}
  \label{fig:valid2}
\end{figure}

The access amplification (total number of memory accesses divided by the demand requests) can be significant in DRAM caches, especially in the case of writes. 
We compare the access amplification of our model with that of the actual hardware.
We use the access amplification numbers presented by Hildebrand et al.~\cite{hildebrand2021case} on an Intel Cascade Lake system.
We calculate the access amplification values for \textit{UDCC} by dividing the effective bandwidth (seen by the LLC) by the sum of the average bandwidth of DRAM and NVRAM devices for a particular run.
The comparison of the two access amplification values is shown in Figure~\ref{fig:valid2}.
Our results match the actual hardware in all cases with only one exception.
The reason for smaller access amplification for write misses is that our implementation slightly optimizes the write requests.
In actual hardware, on a write miss in DRAM cache, the block is first allocated by reading it from NVRAM and then writing it into DRAM. The actual data is then written into the DRAM. We merge these two DRAM writes, and thus, our implementation leads to one less access for a write miss compared to the actual hardware.

Finally, we manually stepped through the operations of different kinds of accesses (reads/writes and hits/misses) to ensure they behaved as expected. 

\section{Methodology}
\label{sec:simMethod}

For all case studies except real workloads under full system simulation, we used \gem{}'s traffic generators instead of execution-based workloads.
Using traffic generators allows us to explore the behavior of the DRAM cache design more directly and more clearly understand the fundamental trade-offs in its design.
We use \gem{}'s traffic generator to generate two patterns of addresses: linear or random.
If not specified, the tests are run with a linear traffic generator.


To implement the DRAM cache we used the DRAM models (``interfaces'') provided by gem5.
We also used the NVM interface provided in gem5 with its default configuration unless otherwise specified (timing parameters are shown in the `Base' column in Table~\ref{tab:nvmtimes}).
We extended \gem to model DDR5 (based on DDR5 data-sheets) and HBM (based on a performance comparison) for our studies.

In the case studies presented below, we are not concerned with the detailed configuration of the \textit{UDCC} (e.g., the size of DRAM cache). 
Instead, we are interested in the study of the behavior of the \textit{UDCC} through specific scenarios 
which enables us to evaluate the best, worst, or in between performances of the system. For this purpose, we have used 
patterns which are either Read-Only (RO), Write-Only (WO), or a combination of reads and writes (70\% reads 30\% writes). 
The other traffic pattern characteristic we varied is the hit (or miss) ratio of the generated pattern. 
This factor was enforced by two different parameters (1) the size of the DRAM cache and (2) the range of addresses requested by the traffic 
generator. In all of the case studies in which the traffic generator was involved, we used a DRAM cache of 16 MB size backed-up 
by NVRAM as the main memory. Unless otherwise specified, we used 
DDR4 to implement the DRAM cache in \textit{UDCC} with a total
buffer size of 256 entries. In order to get 0\%, 
25\%, 50\%, 75\%, and 100\% hit ratios, we set the range of addresses to be 6GB, 64MB, 32MB, 20MB, and 6MB. 
For instance, to study 
the behavior of a write intensive application with a memory foot print larger than the DRAM cache capacity on our proposed model, 
we set the traffic generator to give WO accesses within the range of 6GB for a DRAM cache of 16MB capacity. 
In this way, we were able to test the \textit{UDCC} in a reasonable 
simulation time. 
We used a cache block size of 64B in all tests to match current systems.
All the tests were simulated for 1 second. 



\section{Case Study 1: Impact of Scheduling Policy on DRAM Caches Performance}
\label{sec:sched}

\subsection{Background and Methodology}

The policy to choose which memory request to service has a significant impact on the performance of the DRAM memory system.
In this work we consider two scheduling policies: (i) first-come, first-serve (\textit{FCFS}), and (ii) first-ready, first-come, first-serve (\textit{FRFCFS}).
\textit{FCFS} is the naive policy which processes the memory requests in the order they are received by the memory controller.
While it is simple and easy to implement, \textit{FCFS} adds many row-switching delay penalties, leading to lower bus utilization.
Rixner et al.~\cite{rixner2000memory} proposed \textit{FRFCFS} trying to take advantage of maximizing row-buffer hit rate. \textit{FRFCFS} reorders
the received requests so the ones that hit on the currently-opened row would be serviced earlier than any other requests which map to the other rows that are currently closed.
In heterogeneous memory systems, Wang et al.~\cite{wang2020characterizing} reported that Intel's Cascade Lake's NVRAM interface deploys \textit{FCFS}.
The question that arises here is, in a memory system with two cooperative devices, one as DRAM cache and the other as main memory, with several internal requests buffers to arbitrate from, how important is the choice of scheduling policy?

We extended \textit{UDCC} with \textit{FCFS} and \textit{FRFCFS} scheduling policies to answer this question.
\textit{UDCC} employs the same scheduling policy for both DRAM cache and main memory.
We tested \textit{UDCC} with each
scheduling policy to measure the improvement of bandwidth observed by LLC with \textit{FRFCFS} over \textit{FCFS}.
We have run tests with different hit ratios and different read and write request combinations to test
the sensitivity of bus utilization and the system performance to the scheduling policy.

The results are based on request patterns of 0\% hit ratio and 100\% hit ratio for RO and WO.
We also ran patterns containing 70\% read and 30\% write requests, with 100\%, 75\%, 50\%, 25\% and 0\% hit ratios.

\subsection{Results and Discussions}

\begin{figure}
    \centering
    \includegraphics[scale=0.6]{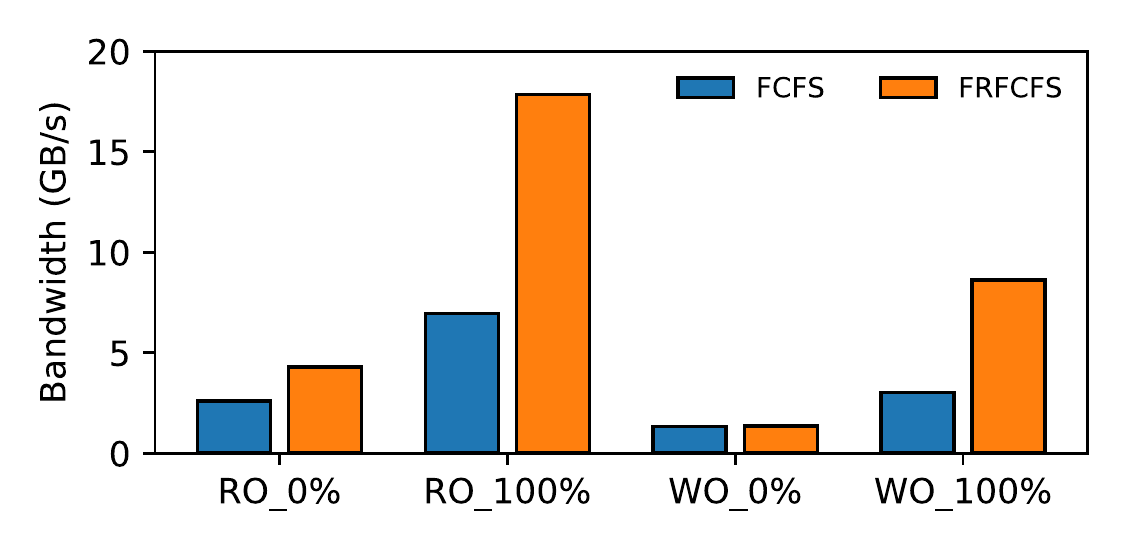}
    \vspace{-1ex}
    \caption{Bandwidth seen by LLC (GB/s). \textit{UDCC} has been tested with a total buffer size of 256 entries, 
    DDR4 DRAM cache, and NVRAM main memory for \textit{FCFS} and
     \textit{FRFCFS} scheduling policies. On the X-axis read-only (RO) and write-only (WO) patterns with 0\% and 100\% hit ratios are
     shown for the two scheduling policies.}
    \label{fig:woroBW}
  \end{figure}

  \begin{figure}
    \centering
    \includegraphics[scale=0.6]{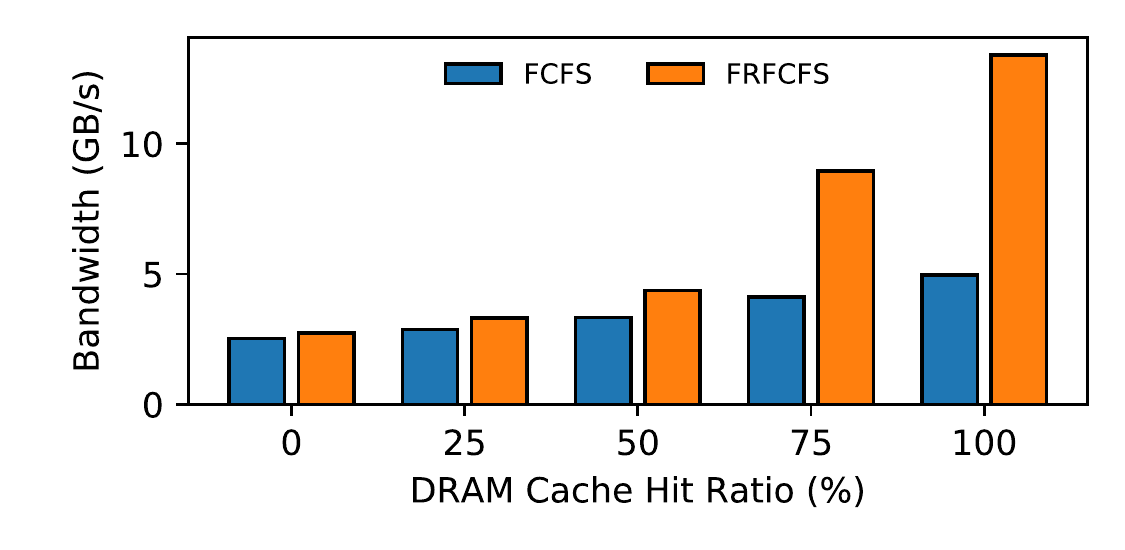}
    \vspace{-1ex}
    \caption{Bandwidth seen by LLC (GB/s). \textit{UDCC} has been tested with a total buffer size of 256 entries, 
    DDR4 DRAM cache, and NVRAM main memory for \textit{FCFS} and \textit{FRFCFS} scheduling policies. \textit{UDCC} has been fed by 70\% read requests and 30\% write requests.
    As shown on the X-axis, different hit ratios (0\%, 25\%, 50\%, 75\%, 100\%) have been applied to the request patterns.}
    \label{fig:r70BW}
  \end{figure}


\begin{figure}
  \subfloat[\scriptsize{DRAM}]{
    \includegraphics[width=.45\linewidth, scale=0.5]{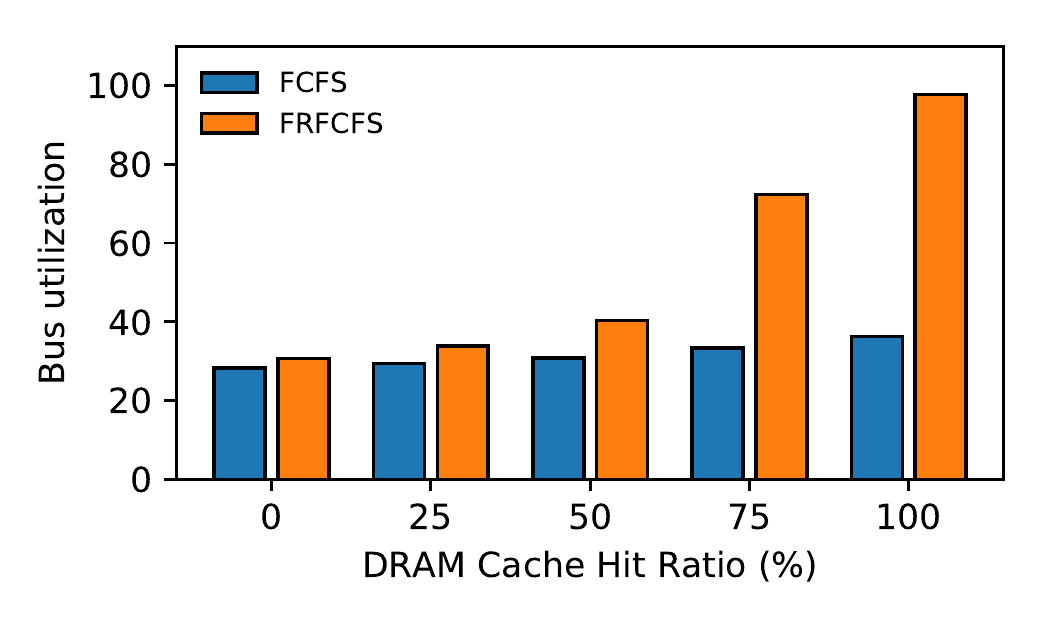}
    \label{fig:dram}
  }
   \subfloat[\scriptsize{NVRAM}]{
    \includegraphics[width=.45\linewidth, scale=0.5]{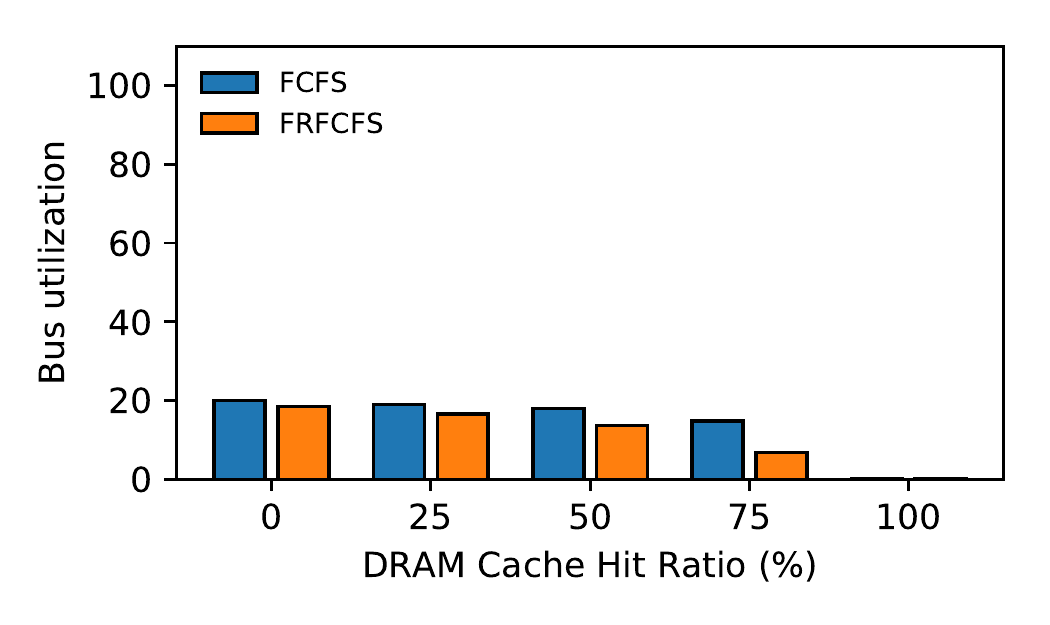}
    \label{fig:nvm}
  }
    \caption{Bus utilization percentage of DRAM and NVRAM. \textit{UDCC} has been tested with a total buffer size of 256 entries, 
    DDR4 DRAM cache, and NVRAM main memory for \textit{FCFS} and \textit{FRFCFS} scheduling policies. \textit{UDCC} has been fed by 70\% read requests and 30\% write requests.
     As shown on the X-axis, different hit ratios (0\%, 25\%, 50\%, 75\%, 100\%) have been applied to the request patterns.}
    \label{fig:r70BusUtil}
  \end{figure}


Figure \ref{fig:woroBW} compares the observed bandwidth at the LLC, while running requests patterns which are RO and WO with 100\% and 0\% hit ratio DRAM cache.
For RO 100\% hit ratio and WO 100\% hit ratio, \textit{FRFCFS} achieved higher bandwidth than \textit{FCFS} by
2.56x and 2.85x, respectively. For RO 0\% hit ratio and WO 0\% hit ratio the observed bandwidth with \textit{FRFCFS} compared to \textit{FCFS} are
1.65x and 1x.

From this data, we conclude that the scheduling policy is more important for workloads which are more likely to saturate the DRAM bandwidth and are not limited by NVRAM bandwidth.
For instance, in the WO with 0\% hit ratio case, the performance is completely limited by the NVRAM device bandwidth and the scheduling policy does not affect the performance at all.

  To look at a slightly more realistic situation, Figure~\ref{fig:r70BW} shows the observed bandwidth by LLC while running a pattern consisting of 70\% read, and
  30\% write accesses and the hit ratio increases on the X-axis with 25\% steps. As the hit ratio increases, the observed bandwidth increases
  for both \textit{FRFCFS} and \textit{FCFS} because more hits results in more DRAM accesses which is higher bandwidth.
  Figure~\ref{fig:r70BW} also shows that as the hit ratio increases, the improvement of bandwidth by \textit{FRFCFS} over \textit{FCFS}, increases.


To better understand the performance improvement of the \textit{FRFCFC} policy 
Figure~\ref{fig:r70BusUtil} shows 
the bus utilization percentage for 70\% read 
and 30\% write case, for both DRAM and NVRAM devices.

  In Figure \ref{fig:dram}, 
  DRAM bus utilization of \textit{FRFCFS} is
  higher than FCFS in all hit ratios. 
  Besides, as the hit ratio increases, the improvement of
  DRAM bus utilization by \textit{FRFCFS} compared to \textit{FCFS} also increases.
  In Figure \ref{fig:nvm}, the NVRAM bus utilization decreases for both \textit{FRFCFS} and \textit{FCFS}, as the hit ratio increases 
  since there are less misses to be handled by NVRAM.
  Overall, DRAM interface bus utilization has benefited more from \textit{FRFCFS} (compared to \textit{FCFS}), than NVRAM. Moreover,
  where the hit ratio is higher, the improvement of bus utilization by \textit{FRFCFS} over \textit{FCFS} is higher.

  Based on the proposed DRAM cache model, there are several internal buffers at the controller per interface for reads and writes. The fair
  arbitration of requests in each of these buffers can highly impact the utilization of the resources, affecting the performance of the DRAM
  cache. 
  We showed
  \textit{FRFCFS} can achieve up-to 2.85x bandwidth improvement over \textit{FCFS} (in WO with 100\% hit ratio) and that the improved DRAM bus utilization is the main factor contributing to this improvement.
  Overall, even though implementation of \textit{FRFCFS} would be more
  costly than \textit{FCFS} due to more complex and associative circuits, the performance gap between these two policies is significant.


\section{Case Study 2: Performance Analysis of Different DRAM Technologies as DRAM Cache}

\begin{figure}
  \subfloat[\scriptsize{Maximum bandwidth achieved}]{
    \includegraphics[width=.48\linewidth, scale=0.5]{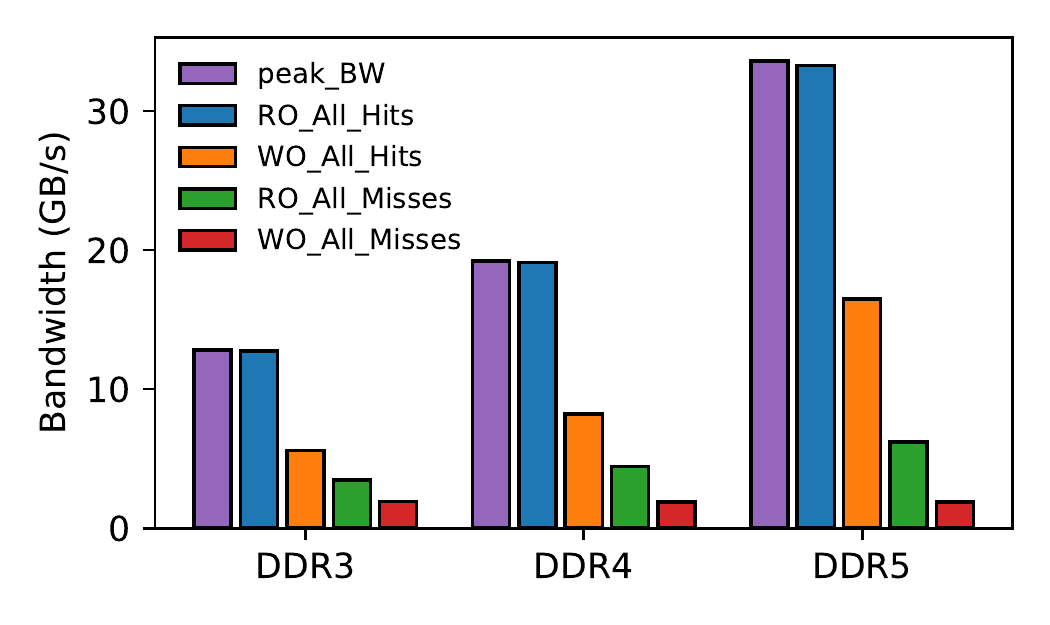}
    \label{fig:bws}
  }
  \subfloat[\scriptsize{Buffer size for maximum bandwidth}]{
    \includegraphics[width=.48\linewidth, scale=0.5]{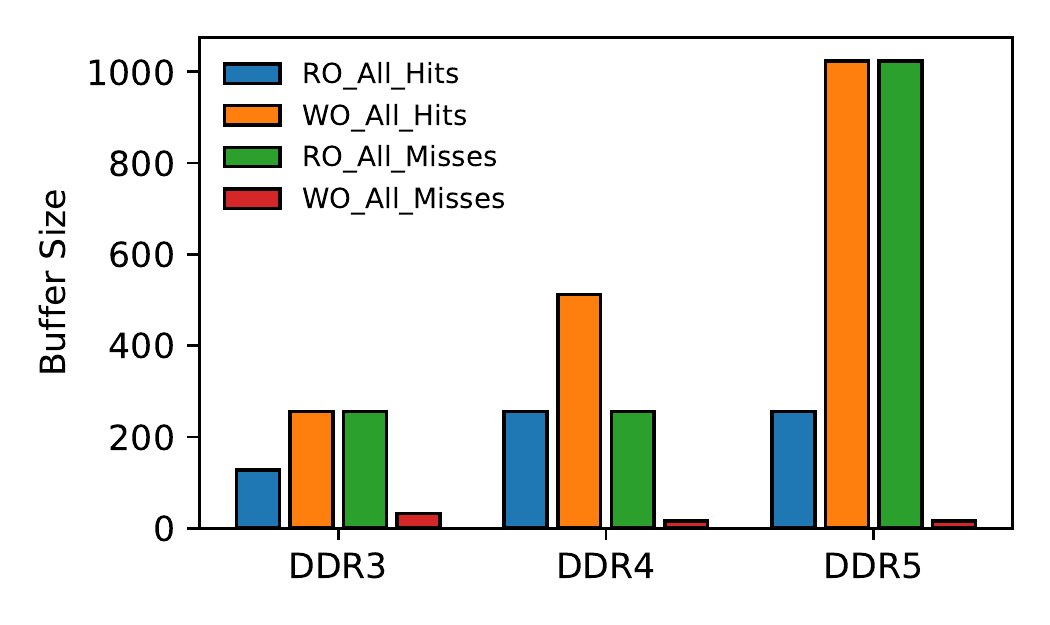}
    \label{fig:orb}
  }
    \caption{Buffer size needed to achieve the maximum bandwidth seen by LLC for 
    \textbf{DDR3}, \textbf{DDR4}, and \textbf{DDR5} as DRAM cache. The traffic patterns 
    shown in the figure include read-only (RO) and write-only (WO), 
    each having 100\% and 0\% hit ratio.
    }
    \label{fig:orb_bw}
  \end{figure}

\subsection{Background and Methodology}

Different commercial products have adapted different DRAM technologies for their DRAM caches (e.g., HBM in Sapphire Rapids and DDR4 in Cascade Lake).
These technologies have different characteristics (e.g., peak bandwidth) that gives different applicability to each.
In this
study we want to answer these questions: how many total buffers are required for each DRAM technology to fully utilize it as DRAM cache?
And, what is the peak performance of each device when the hit ratio and the percentage of read-write in access patterns change?

To address these questions, we configured \textit{UDCC} to use DDR3, DDR4, DDR5 and HBM models implemented for \gem, as DRAM cache.
The theoretical peak bandwidth of DDR3, DDR4, DDR5 and HBM are 12.8 GB/s, 19.2 GB/s, 33.6 GB/s, and 256 GB/s, respectively.
The results are based on request patterns of 
RO 100\% hit ratio, RO 100\% miss ratio, WO 100\% hit ratio, and WO 100\% miss ratio.
We ran these patterns across all DRAM technologies for buffer sizes from 2 to 1024 by powers of 2.
For each
case we looked for the buffer size where the observed bandwidth by LLC reached to its maximum state and
would not get improved by increasing the size of the buffers.
To separate out the total number of buffers from the specific micro-architectural decisions (e.g., how to size each buffer), we only constrained the ``outstanding request buffer'' and allowed all of the internal buffers to be any size.

\subsection{Results and Discussion}

Figure \ref{fig:bws} shows the maximum achieved bandwidth for the DRAM cache, using DDR3, DDR4 and DDR5,
for the case of RO 100\% hit ratio, RO 100\% miss ratio, WO 100\% hit ratio, and WO 100\% miss ratio.
Moreover, Figure \ref{fig:orb} shows the amount of total buffers required in order to reach to the maximum bandwidth shown
in Figure \ref{fig:bws} for each case.

In RO 100\% hit ratio access pattern, each request requires only a single access to fetch the data along with tag and metadata from DRAM.
Thus, it is expected to achieve to a bandwidth very close to the theoretical peak bandwidth, and Figure~\ref{fig:bws} shows this is the case.
Moreover,
the Figure \ref{fig:orb} shows that DDR3, DDR4, and DDR5 have reached to this bandwidth at buffer size of 128, 256, 256, respectively. Thus, increasing the buffer size would not help for RO 100\% hit traffic pattern.

In WO 100\% hit ratio access pattern, each request requires two accesses to fetch tag and metadata from DRAM and then writing the data to the DRAM.
The peak bandwidth in this case for each device is lower than the theoretical peak by about a factor of two (5.58 GB/s, 8.2 GB/s, and 16.47 GB/s for DDR3, DDR4, and DDR5, respectively).
Since these requests require two DRAM accesses, the latency of each request is higher and more buffering is required to reach the peak bandwidth.
Specifically, Figure~\ref{fig:orb} shows that DDR3, DDR4, and DDR5 have reached to this bandwidth at buffer size of 256, 512, 1024~\footnote{DDR5 shows bandwidth improvement of 8\% going from 512 to 1024 entries}, respectively.
Comparing this case with RO 100\% hit ratio shows this case has gotten some bandwidth improvement from increasing the buffer size.

In RO 100\% miss ratio access pattern, each request requires three accesses: one for fetching tag and metadata from DRAM, then fetching
the line from NVRAM, and finally writing the data to the DRAM.
The peak bandwidth in this case for each device is 3.47 GB/s, 4.45 GB/s, and 6.21 GB/s for DDR3, DDR4, and DDR5, respectively.
Figure \ref{fig:orb} shows that DDR3, DDR4, and DDR5 have reached to this bandwidth at buffer size of 256, 256, 1024, respectively.
It suggests DDR5 has gotten some bandwidth improvement from increasing the buffer size. 
However, the bandwidth improvement from 512 to 1024 buffer entries is 3.8\% (and 2\% for 256 to 512 buffer size change).

Finally, for WO 100\% miss ratio access pattern, each request requires four accesses: one for fetching tag and metadata from DRAM, then fetching
the line from NVRAM, writing the data to the DRAM, and a write to NVRAM for dirty line write back. 
The peak bandwidth in this case is 1.91 GB/s, 1.90 GB/s, and 1.90 GB/s for DDR3, DDR4, and DDR5, respectively.
The Figure \ref{fig:orb} shows that DDR3, DDR4, and DDR5 have reached to this bandwidth at buffer size of 16, 16, 32, respectively.
Comparing this case with the previous cases shows that, in this case the bandwidth gets saturated in smaller
buffer sizes than the other cases.



The main takeaway from this case study is that the buffer size required to achieve peak bandwidth largely depends on
the composition of memory traffic. Secondly, the DRAM cache controller might need a large number of buffers to reach the peak
bandwidth, primarily if the device can provide a large bandwidth (e.g., over 1000 in the case of DDR5).
More memory controller buffers are needed because of the increased latency of accesses since each write request and read miss request requires multiple accesses to the memory devices.


\subsection{HBM}

We separate out HBM from the discussion of other DDR memory technologies as its memory controller design may be different from the \textit{UDCC} design described in this paper and used in the Intel's Cascade Lake.
For instance, in Intel's Knight's Landing, the DRAM cache was implemented with a four-state coherence protocol and the HBM interfaces were physically separated from the DRAM interfaces through the on-chip network~\cite{sodani2015knights}.

\begin{figure}
  \centering
  \includegraphics[scale=0.6]{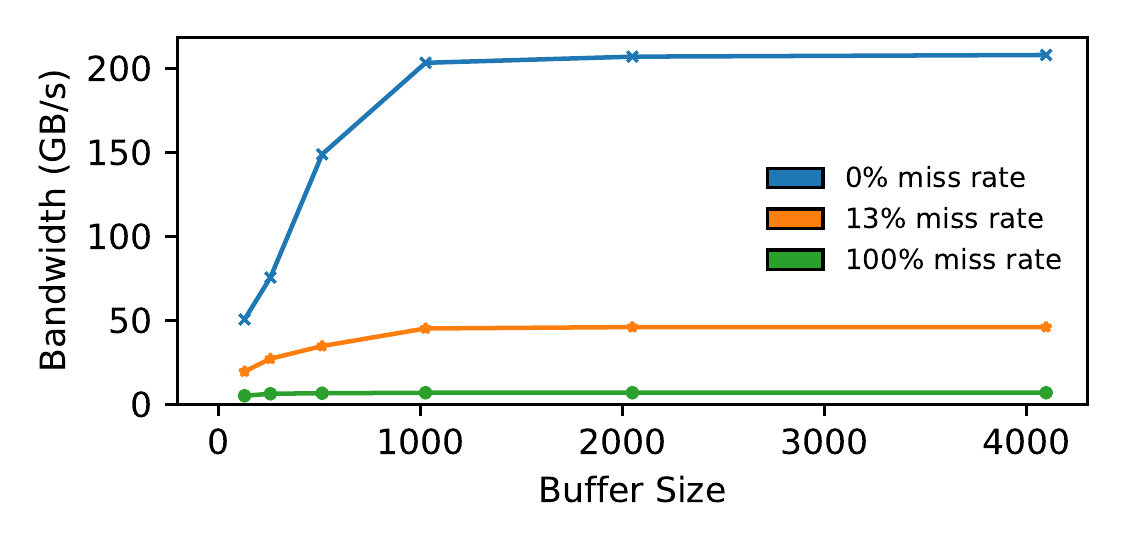}
  \vspace{-1ex}
  \caption{Buffer size impact on read bandwidth for an HBM based DRAM cache (theoretical peak bandwidth : 256 GB/s).}
  \label{fig:hbm}
\end{figure}

We modeled an HBM interface in \gem{} such that it can provide the theoretical peak bandwidth of 256 GB/s in a single memory channel (equivalent to the sum of all pseudo channels in an HBM2 device).
Figure~\ref{fig:hbm} shows the bandwidth that is achieved for different total buffer sizes, for two cases: 1) when all accesses hit in the DRAM cache, and 2) when all accesses miss in the DRAM cache.
In case of all hits, we see close to the maximum possible bandwidth with an 2048 buffers.
In case of all misses, the achievable bandwidth is limited by the main memory used in the experiment (a NVRAM device) and does not get better after buffer size of 1024 entries.

This data implies that the buffer sizes required for HBM are not that much higher than for high-performance DDR DRAMs (e.g., DDR5).
The main reason that HBM does not require significantly more buffering is because even for a small miss rate (13\%), we are limited by the NVRAM's performance and saturate bandwidth at much less than the theoretical peak as shown in Figure~\ref{fig:hbm}.
However, if the HBM cache was backed by higher performance memory (e.g., DDR4 or DDR5 DRAM) the total number of buffers to achieve the maximum performance would likely be much higher.

\section{Case Study 3: Performance Analysis of DRAM Caches While Backed-up by Different Main Memory Models}

\subsection{Background and Methodology}

\begin{table}
	\centering
       \caption{NVRAM Interface Timing Parameters}
      \begin{tabular}{|p{2.2cm}|p{1.32cm}|p{1.50cm}|p{1.40cm}|}
      \Xhline{2\arrayrulewidth}
      \textbf{Parameter} & \textbf{Slow} & \textbf{Base} & \textbf{Fast} \\ \Xhline{2\arrayrulewidth}
      \textbf{tREAD} & 300ns & 150ns & 75ns \\ \hline
      \textbf{tWRITE} & 1000ns & 500ns & 250ns \\ \hline
      \textbf{tSEND} & 28.32ns & 14.16ns & 7.08ns \\ \hline
      \textbf{tBURST} & 6.664ns & 3.332ns & 1.666ns \\ \Xhline{2\arrayrulewidth}
    \end{tabular}
    \label{tab:nvmtimes}
\end{table}


In this section, we focus on the question how much does the performance (such as latency) of the main memory affect the performance of the memory system as observed by the LLC in DRAM cache-based systems? 
We extended the baseline
NVRAM interface model of gem5 to implement a faster NVRAM and slower NVRAM compared to the baseline performance.
The baseline, fast, and slow models of gem5 NVRAM provide 19.2 GB/s, 38.4 GB/s, and 9.6 GB/s bandwidth, respectively.
Table \ref{tab:nvmtimes} shows the timing constraints of all three cases.
The results are based on
an access pattern with a high miss ratio and large number of dirty line evictions from DRAM cache,
to highly challenge the performance of the main memory for both read and write accesses to it.
For this purpose, we used a WO with 100\% miss ratio access pattern which generates dirty lines along the way,
requiring write-backs to the main memory. This access pattern highly engages the NVRAM during miss 
handling to fetch the missed lines
and write-back of dirty lines, enabling us to evaluate the performance of the system 
in all three cases of slow, fast, and baseline NVRAMs. Moreover, we have tested them for a pattern 
consisting of RO 100\% miss ratio, since this case also requires interaction with NVRAM (for fetching 
the missed line). In both patterns, the results are based on a total buffer size of 512 entries.

\subsection{Results and Discussion}

\begin{figure}
  \centering
  \includegraphics[scale=0.5]{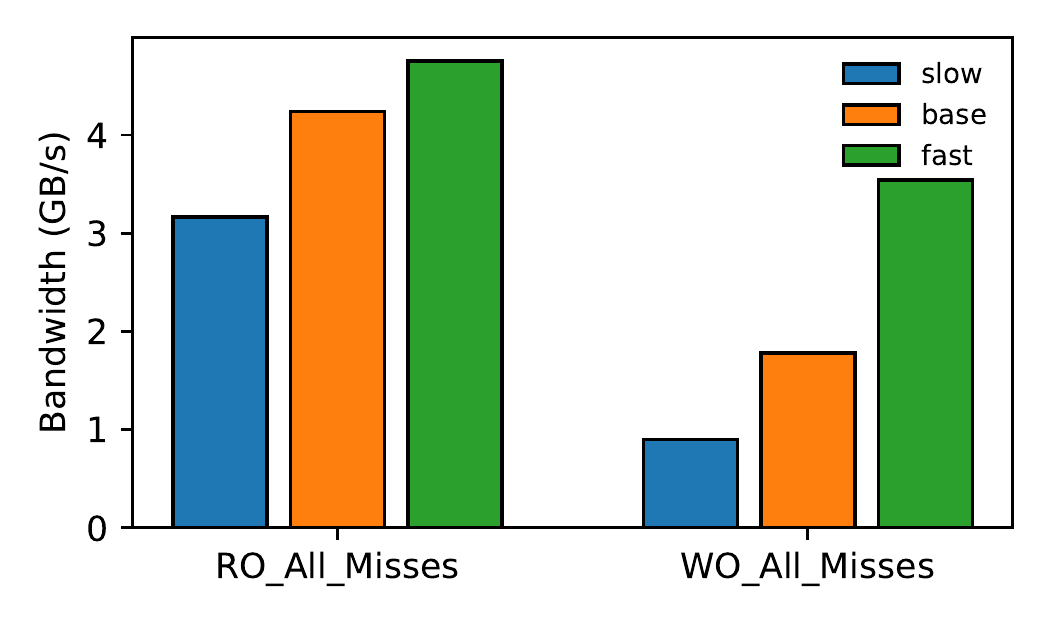}
  \vspace{-1ex}
  \caption{Observed bandwidth by the LLC for read-only (RO) 100\% miss ratio and write-only (WO) 100\% miss ratio.}
  \label{fig:bw-wo100M-cs4}
\end{figure}


First we investigate the effect of the different NVRAMs from the UDCC-external point of view.
Figure \ref{fig:bw-wo100M-cs4} compares the bandwidth seen by the LLC for three
different NVRAMs, slow, baseline, and fast, for RO 100\% miss ratio and WO 100\% miss ratio 
requests patterns. Note that NVRAM has a 
dedicated write buffer in \textit{UDCC} whose size is enforced by the NVRAM interface (128 entries in gem5 NVRAM models). 
Our resuts showed, the average queuing latency for this buffer is 76.19~$\mu$s,
38.49~$\mu$s, and 19.52~$\mu$s for slow, baseline, and fast NVRAMs, respectively.
In other words, the write queuing latency at the NVRAM interface, gets shorter
once the speed of NVRAM increases, as expected. 
In the WO 100\% miss ratio, the highest achieved bandwidth is 0.96 GB/s, 1.91 GB/s, and 3.8 GB/s for slow, baseline and fast NVRAMs, 
respectively. This interprets to 49.7\% bandwidth degradation once the
slow NVRAM was used as the main memory, and 98.9\% bandwidth improvement when the fast NVRAM
was used as the main memory, compared to the baseline NVRAM. Note that in this request pattern 
there are two NVRAM accesses (one read and one write).

In the RO 100\% miss ratio, the highest achieved bandwidth is 3.39 GB/s, 4.55 GB/s, and 5.1 GB/s for slow, baseline and fast NVRAMs, 
respectively. This interprets to 25.49\% bandwidth degradation once the
slow NVRAM was used as the main memory, and 12.08\% bandwidth improvement when the fast NVRAM
was used as the main memory, compared to the baseline NVRAM. Note that in this request pattern 
there is only one NVRAM (read) access.

These results suggest that if an access pattern requires more interaction with NVRAM (e.g., 
where the DRAM cache miss ratio is higher, or when there are more dirty line evictions), 
the improvement of bandwidth observed by LLC with faster NVRAM is higher, for a system consisting of DRAM cache. 
Moreover, these results are based on a total buffer size of 512 entries as our further investigations showed that 
none of the devices gains more bandwidth by providing larger buffer size. Thus, even though slowing down 
the NVRAM device hurts performance, it does not affect the microarchitectural details of the controller.

\section{Case Study 4: Evaluating DRAM Caches for Real Applications}
\label{sec:realBench}

In this case study, we performed an evaluation of DRAM caches for real-world applications, i.e., NPB~\cite{bailey1991parallel} and GAPBS~\cite{beamer2015gap}.
The DRAM caches are expected to perform similar to a DRAM main memory if the working set of the workload fits in the cache.
Therefore, a case of particular interest for us is to evaluate the performance of the DRAM cache-based system when the workload does not fit in the DRAM cache.
To accomplish the previously mentioned goal, we model a scaled-down system with 64MB DRAM cache and run NPB and GAPBS workloads for one second of simulation time.

\begin{figure}
  \subfloat[\scriptsize{Performance}]{
    \includegraphics[width=.55\linewidth, scale=0.6]{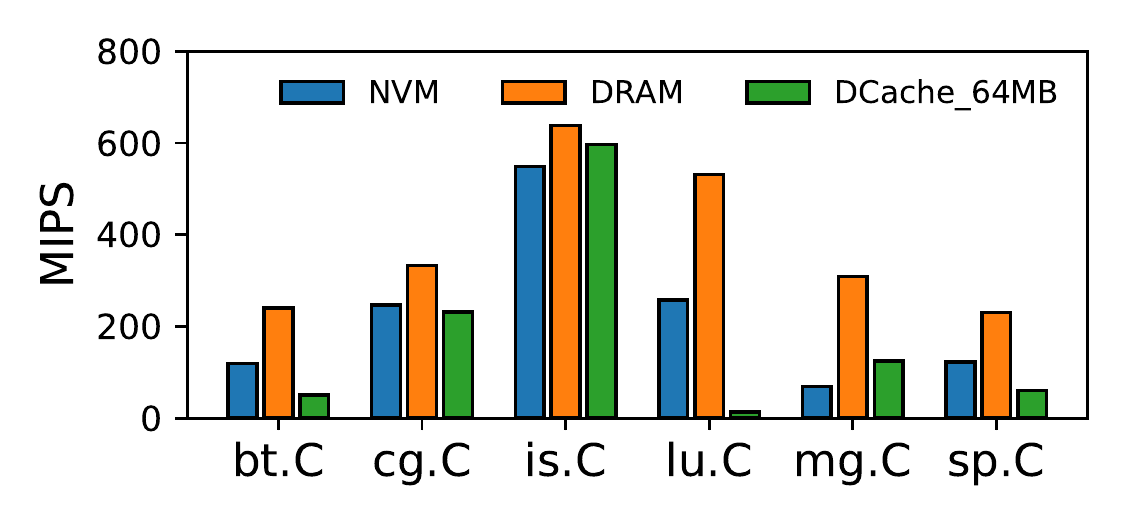}
    \label{fig:npb}
  }
   \subfloat[\scriptsize{Cache miss rate}]{
    \includegraphics[width=.45\linewidth, scale=0.5]{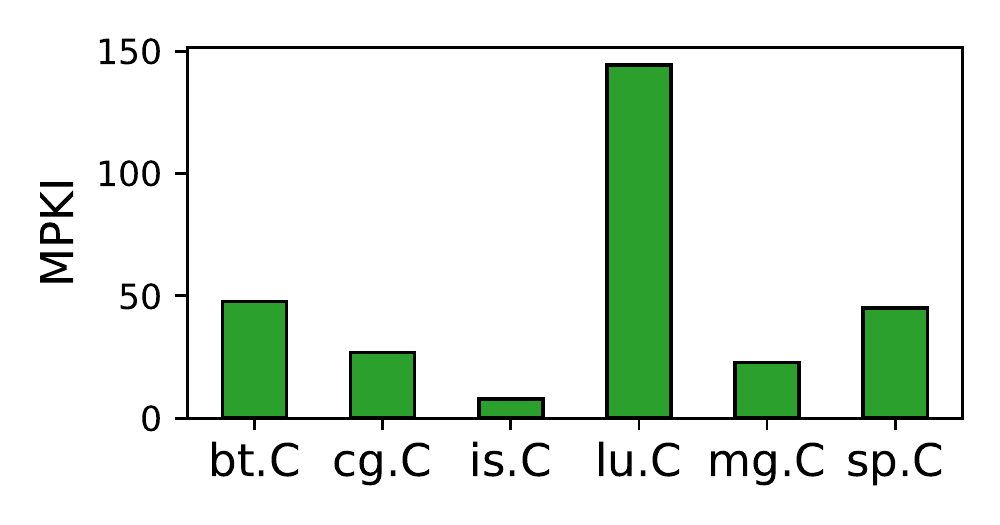}
    \label{fig:npbmiss}
  }
    \caption{NAS Parallel Benchmarks on DRAM Cache}
    \label{fig:npbRes}
  \end{figure}

We run all workloads in three different configurations.
The first two configurations (\textit{NVRAM, DRAM}) model a system without a DRAM cache and the main memory as NVRAM or DRAM.
The third configuration (\textit{DCache\_64MB}) uses a 64MB DRAM cache and NVRAM as main memory.
Figure~\ref{fig:npb} shows a million instructions per second (MIPS) values for NPB in three configurations.
In most cases, \textit{DCache\_64MB} performs the worst, with the most prominent performance degradation for \textit{lu.C} and \textit{bt.C}.
The only exception is \textit{is.C}, where \textit{DCache\_64MB} performs better than \textit{NVRAM}.
The performance of \textit{DCache\_64MB} correlates with the DRAM cache misses per thousand instructions (MPKI) values shown in Figure~\ref{fig:npbmiss}.
For example, \textit{is.C} shows the smallest and \textit{lu.C} shows the largest MPKI values.

\begin{figure}
  \subfloat[\scriptsize{Performance}]{
    \includegraphics[width=.55\linewidth, scale=0.6]{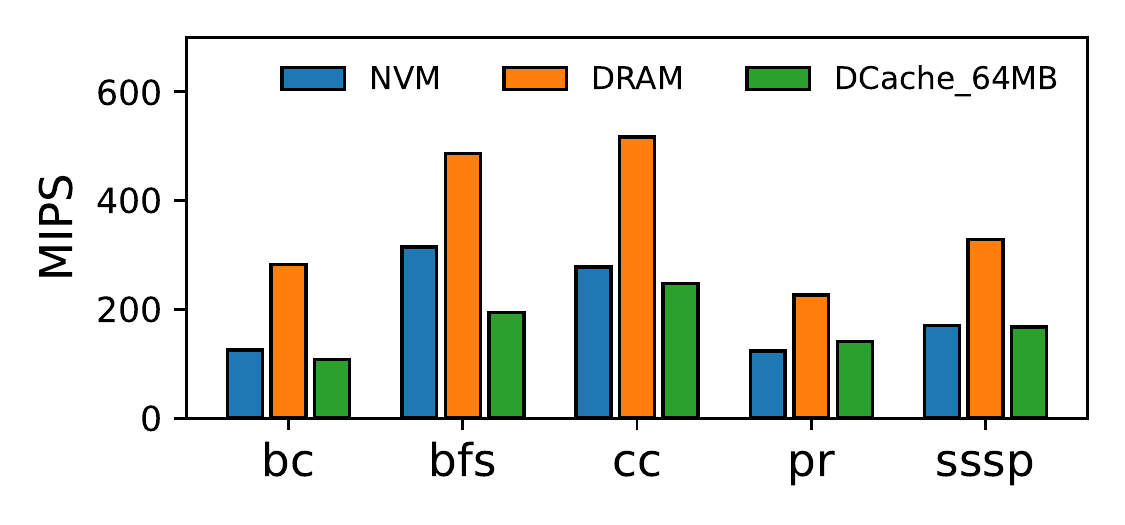}
    \label{fig:gapbs}
  }
   \subfloat[\scriptsize{Cache miss rate}]{
    \includegraphics[width=.45\linewidth, scale=0.5]{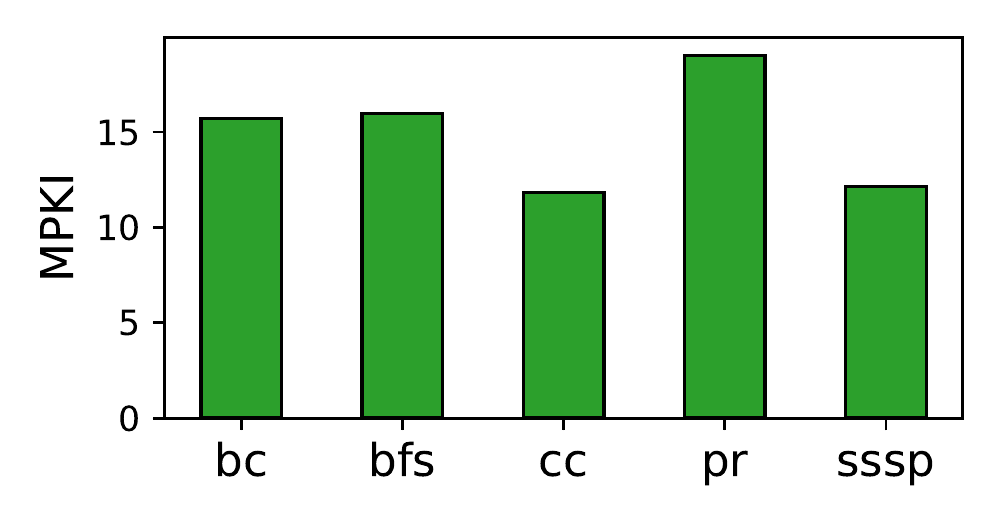}
    \label{fig:gapbmiss}
  }
    \caption{GAPBS on DRAM Cache}
    \label{fig:gapbsRes}
  \end{figure}

Figure~\ref{fig:gapbs} shows a million instructions per second (MIPS) values for GAPBS in the previously mentioned three configurations.
Figure~\ref{fig:gapbmiss} shows the associated MPKI values.
The graph workloads on a DRAM cache perform mostly similarly to NVRAM alone in our simulation runs.
The only exception is \textit{bfs} which shows a significantly lower MIPS value than \textit{NVRAM}.
The DRAM cache MPKI value for \textit{bfs} alone does not explain the reason for this behavior.
\textit{bfs} has the highest fraction of writes than reads in the requests that hit the DRAM cache (29\% more writes than reads in case of \textit{bfs} in contrast to 36\% less writes than reads for the rest of the workloads)

Since a write hit also leads to access amplification (tag read before a write hit), the impact of this amplification is then seen in the performance degradation with a DRAM cache.
We conclude from this observation that the extra DRAM accesses (for tag reads) can also impact the workload's performance on a DRAM cache.

\section{Case Study 5: NVRAM Write Wear Leveling Effect}

\subsection{Background and Methodology}
Non-volatile storage and memory devices, such as Phase Change Memories (PCM) used as NVRAM
and flash memory, are known to have a limited write endurance.
Write endurance is defined as the number of writes to a block in the device, before it becomes
unreliable. One of the techniques used for such devices to prolong their lifetime, is
wear leveling. Usually NVRAM wear-leveling technique tries to evenly distribute wear-out by moving data from one
highly-written location to a less worn-out location. Wang et al.~\cite{wang2020characterizing} measured the frequency and latency overhead
of data migration due to wear leveling in NVRAMs through probing and profiling. They reported a long tail latency of 60~$\mu$s
for almost every 14000 writes to each 256B region of NVRAM. Wear leveling can affect the performance of DRAM caches. 
This effect is more noticeable while running write-intensive
workloads which their memory footprint is larger than the DRAM cache capacity. In such situation, many dirty lines eviction will happen
that are needed to be written back to NVRAM; thus, an overall increase on NVRAM writes can be expected. As a result, frequent data migration for
wear-leveling (with a long tail latency) will be done by NVRAM interface and a performance degradation
can be expected in this case.

In this section we investigate the effect of this long latency on DRAM cache performance. We extended NVRAM interface model of
gem5 so for every 14000 write accesses the interface adds 60~$\mu$s extra latency, delaying the service time of the next request.
The access pattern was set to be all writes and 100\% miss and 8x bigger than the DRAM cache size. This will
increase the number of write-backs of dirty lines
from DRAM cache to the NVRAM, to pressure the NVRAM write buffer and see the effect of wear leveling delay on the overall system performance.

\subsection{Results and Discussion}

\begin{table}
	\centering
       \caption{Maximum write bandwidth (and average NVM write latency) achieved with and without wear-leveling}
      \begin{tabular}{|p{2.3cm}|p{2.45cm}|p{2.75cm}|}
      \Xhline{2\arrayrulewidth}
      \textbf{Parameter} & \textbf{With wear-leveling} & \textbf{Without wear-leveling} \\ \Xhline{2\arrayrulewidth}
      \textbf{Bandwidth} & 1.77 GB/s & 1.92 GB/s \\ \hline
      \textbf{NVM write latency} & 42.84~$\mu$s & 39.71~$\mu$s \\ \Xhline{2\arrayrulewidth}
    \end{tabular}
    \label{tab:nvmwear}
\end{table}

Table \ref{tab:nvmwear} compares the overall bandwidth seen by LLC in two cases; with wear leveling and
without wear leveling. 
Without wear leveling the peak bandwidth
is 1.92~GB/s while it drops to~1.77 GB/s when wear leveling is activated. These results show a 7.8\% performance degradation which directly comes
from wear leveling overhead.
Table \ref{tab:nvmwear} also shows the average queuing latency measured for NVRAM write buffer. This latency is 42.84~$\mu$s for
NVRAM-with-wear-leveling, and 39.71~$\mu$s for NVRAM-without-wear-leveling.
Thus, the 60~$\mu$s latency considered for data migration during wear leveling, has caused 7.8\% latency overhead on NVRAM write buffer as well.
This 7.8\% overhead is larger than expected given the rarity of wear leveling events showing that these rare events have an outsized impact when the system is configured with a DRAM cache.



\section{Related work}

Most of the prior research work in DRAM cache organization do not provide detailed methodologies required to model a DRAM cache during the simulation. In terms of modeling of a 
DRAM cache, Gulur et al.~\cite{gulur2015comprehensive} presented an analytical performance model of DRAM cache for in-SRAM and in-DRAM tag storage organizations. 
Their model considers parameters such as DRAM Cache's and off-chip memory's timing values, cache block size, tag cache/predictor hit rate and workload characteristic, 
to estimate average miss penalty and bandwidth seen by the last level on chip SRAM cache (LLSC). Their work is based on a prior model called ANATOMY~\cite{gulur2014anatomy} which is a trace-based analytical model estimating 
key workload characteristics like arrival rate, row-buffer hit rate, and request spread, that are used as inputs to the network-like queuing model to statistically estimate memory performance. 
Even though this work accounts for the timing constraints of DRAM for the cache management, it is agnostic of the microarchitectural and timing constraints of main memory technologies cooperating with DRAM cache, 
and still leaves a gap for a full system DRAM cache simulation for detailed architectural analysis.



Wang et al.~\cite{wang2020characterizing} presented VANS, a cycle-level NVRAM simulator which, models the microarchitectural details of Optane DIMMs.
However, their simulation model does not support using an NVRAM device as a backing memory of a DRAM cache.
VANS can be integrated with other simulators like \gem{}.
In our work, we rely on \gem{}'s default NVM model, but we plan to use the VANS detailed NVRAM model in the future.

\section{Conclusion}

In this work, we described our detailed cycle-level DRAM cache
simulation model implemented and validated on \gem{}, which can enable performing a design space exploration in
DRAM cache research.
This new tool can be used to explore new unified DRAM cache and memory controller designs as part of an agile and full-system simulation platforms.
The tool we presented in this work
can enable many interesting research work in the domain of heterogeneous memory systems.
For instance, using \textit{UDCC} we can address questions such as what is
the efficient data movement or data placement in systems composed of fast and slow memories.
Since, our tool provides full system simulation platform, also it can address the hardware and software co-design
ideas to explore design space of heterogeneous memories.

Moreover, \gem{} is highly modular and allows composing a simulated system based on a variety of components.
\textit{UDCC} can enable experimenting with different and new memory device models which might have features to be a better fit to be used as a cache to a backing memory.


\bibliographystyle{IEEEtran}
\bibliography{IEEEabrv,references}

\end{document}